\documentclass[12pt]{article}
\usepackage{amsmath}
\usepackage{graphicx}
\usepackage{enumerate}
\usepackage{enumitem}
\usepackage{hyperref}
\usepackage{listings}
 \usepackage{setspace}
 \usepackage{dblfloatfix}

\usepackage[svgnames]{xcolor}
\usepackage{listings}

\usepackage{url}

\usepackage{geometry}    
\usepackage{float}
\usepackage{wrapfig}
\usepackage{sidecap}								
\usepackage{amssymb}
\usepackage{amssymb} %maths
\usepackage{amsmath} %maths
\usepackage{lipsum}
\usepackage{array} 
\usepackage{caption}
\usepackage{subcaption}
\usepackage{booktabs,caption}
\usepackage[flushleft]{threeparttable}
\usepackage{setspace}

\newcommand{\blind}{1}

\graphicspath{ {./images/} }
\begin{document}

\def\spacingset#1{\renewcommand{\baselinestretch}%
{#1}\small\normalsize} \spacingset{0}

%%%%%%%%%%%%%%%%%%%%%%%%%%%%%%%%%%%%%%%%%%%%%%%%%%%%%%%%%%%%%%%%%%%%%%%%%%%%%%

\if1\blind
{
  \title{\bf Revisiting the Conceptualization of Multiple Linear Regression}
  \author{Grayson L. Baird  \\
Brown University and Rhode Island Hospital \\
    and \\\
    Stephen L. Bieber \\\
University of Wyoming}
  \maketitle
} \fi

\if0\blind
{
  \bigskip
  \bigskip
  \bigskip
  \begin{center}
    {\LARGE\bf Revisiting the Conceptualization of Multiple Linear Regression}
\end{center}
  \medskip
} \fi

\bigskip
\begin{abstract}
The problem known as multicolinearity has long been recognized to fundamentally and negatively influence multiple regression. This paper does not intend to either propose a numerical assessment of the degree to which this problem exists within any data set or a solution to the problem itself. Rather, it is our intent to illustrate the potentially serious ramifications multicolinearity has on the traditional development of the multiple linear regression (MLR) model and its associated statistics using established equations, Venn diagrams, and real data.

\end{abstract}

\noindent%
\vfill

\newpage
\spacingset{1.9}
\section{Background}
\label{sec:intro}

The problem known as multicolinearity has long been recognized to fundamentally and negatively influence multiple regression. It is not the intent of this paper to propose either a numerical assessment of the degree to which this problem exists within any dataset or to propose a solution to the problem itself. Rather, it is our intent to illustrate the potentially serious ramifications multicolinearity has on the traditional development of the multiple linear regression (MLR) model and its associated statistics.

The heart of the problem in the traditional development of MLR seems to stem from the basic analysis of variance identity, in which the variation within the dependent variable (Y) can be expressed as SS(total) = SS(model) + SS(residual). Although this logically leads in regression to 

   \begin{equation} 
    SS(total) = SS(regression) + SS(residual)
    \end{equation}, as seen in Woolf (1951), Baird and Bieber (2016, 2020), and below, equation (1) does not hold for multiple regression problems in which the predictors are correlated. The resultant impact of this violation can be rather extensive and serious. As an aside, as in analysis of variance situations, equation (1) does hold in regression contexts in which some orthogonalization of the variable set is used (Yates, 1938). However, for any MLR model in which any or all of the predictors (e.g., X\textsubscript{1}, X\textsubscript{2}) are correlated with each other and Y (Figure 1), it will never hold. The calculation of the sum of squares (regression) = SS(regression) is usually determined either directly as

\begin{equation} SS(regression)=
\sum_{i=1}^{p} {\textup{b}_i} \sum_{i=1}^{p} {\textup{X}_i{Y}} = {\textup{b}'{X}'{Y}}
\end{equation} or indirectly once SS(residual) has been calculated as 

\begin{equation} SS(regression)=
\\SS(total)-SS(residual)
\end{equation}. From this point onward, to avoid confusion, we will only refer to SS(regression) and SS(residual) in 
association with their calculation through equations (1), (2), and (3). The sum of squares associated with a particular regression model will be symbolically referred to as SS(variables included). For instance, the model sum of squares for the MLR model using the variables X\textsubscript{1} and X\textsubscript{2} would be SS(X\textsubscript{1},X\textsubscript{2}).

To facilitate the presentation of the basic principles and for the confirmation of the assertions above, the Dwaine Studios, Inc. data used to present multiple regression in {\textit{Applied Linear Statistical Models (Kutner, Nachtsheim, Neter, and Li; 2005, 5th edition, p. 237)}} will be used here; these data and corresponding R code (R Foundation for Statistical Computing, Vienna, Austria) are presented in Appendix A. However, any dataset and the development of the fundamental statistics associated with MLR from any source would work equally well for this discussion. The dependent variable (Sales) is represented by Y throughout our discussion, and similarly the independent variables (TARGTPOP and DISPOINC) will be represented by X\textsubscript{1} and X\textsubscript{2}. Table 1 presents selected results from the simple regressions and the multiple regression (MLR) using X\textsubscript{1} and X\textsubscript{2} to predict Y.

\begin{SCfigure}
  \centering
  \caption{Relationship between Predictors (X\textsubscript{{1}}), (X\textsubscript{{2}}), and Dependent Variable (Y) that are intercorrelated}
   \includegraphics[scale=.40] %
    {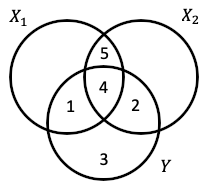}% picture filename
\end{SCfigure}

Using the MLR coefficients presented in Table 1 along with the sums of squares and cross-products provided in Table 2, the elements of equation (1) are

\begin{equation*}      
SS(total)=\\Y'Y= 26,196.21
\footnote{All numerical values in this paper have been rounded-off to 2 decimal places; however, all calculations have been done with full computer precision. In addition, the results are for mean centered data.}
\end{equation*}

\begin{equation*} 
\begin{aligned}
& SS(regression)=b'X'Y=\ {\textup{b}_1}'{\textup{X}_1}'Y+{\textup{b}_2}'{\textup{X}_2}'Y\\
&=(1.452)(12,730.59) + (9.366)(587.04) = 24,015.28 
\end{aligned}
\end{equation*}

\begin{equation*} 
\begin{aligned} 
SS(residual)=SS(total)-SS(regression)=26,196.21-24,015.28 = 2,180.93
\end{aligned}
\end{equation*} These results above can be confirmed in Kutner, Nachtsheim, Neter, and Li (2005) or by application of any statistical computer package. They provide the foundation as a {\textit{fait accompli}}  for the presentation of either equation (2) or (3) as the correct determination of the sum of squares associated with the MLR model. The implication is that SS(regression) = SS(X\textsubscript{1},X\textsubscript{2}) in the MLR situation. 

The question we would like to raise at this point is the following: is the information provided above for equations (1), (2), and (3) actually correct? More specifically, is the calculation of SS(regression) correct, in which it is assumed that SS(regression) = SS(X1, X2)?  In an attempt to answer this question, we would like to digress for a moment and briefly introduce the work presented in Woolf (1951). Woolf’s work is not being presented as the answer or even as an answer to the multicolinearity problem, his work is merely being used to assist us in answering the question above. 
     
In 1951 computers were not readily available to researchers to make the relatively complex calculations associated with multiple regression. To assist researchers, Woolf (1951) wrote his paper on the  {\textit{Computation and Interpretation of Multiple Regressions}}. His second method of calculating the regression coefficients and analysis of variance sum of squares is known often as the determination of Extra Sum of Squares. 

     Very simply, for our two predictor problem, the sum of squares for regression is calculated as 
     
\begin{equation} SS(regression)= SS(X\textsubscript{1}) + SS(X\textsubscript{2}|X\textsubscript{1}) 
\end{equation} where the predictors are ordered in their consideration. In equation (4), X\textsubscript{1} is considered first and then X\textsubscript{2}. The Extra Sum of Squares associated with X\textsubscript{2}, the sum of squares added to the explanation of Y beyond those already established by X\textsubscript{1}, is calculated as the sum of squares associated with the partial predictor (X\textsubscript{2} \textbar X\textsubscript{1}) as SS(X\textsubscript{2}\textbar X\textsubscript{1}) = SS(regression)–SS(X\textsubscript{1}).\footnote{SS(regression) as calculated by equation (2) or (3) is intentional indicated here rather than SS(X\textsubscript{1},X\textsubscript{2}).} This concept is presented in analysis of variance as the sequential determination of the sum of squares known as Type I. From the information provided in Table 1, SS(X\textsubscript{2} \textbar X\textsubscript{1}) = 24,015.28 – 23,371.81 = 643.99. It is worth noting that:

\begin{equation} SS(regression)= SS(X\textsubscript{1}) + SS(X\textsubscript{2}|X\textsubscript{1}) = 23,371.81 + 643.81 = 24,015.28
\end{equation} This result is depicted in Figure 2 where regions 1 and 4 are accounted for by SS(X\textsubscript{1}) and region 2 is accounted for by SS(X\textsubscript{2} \textbar X\textsubscript{1}). 

\begin{SCfigure}
  \centering
  \caption{Variation of Y Explained by the Orthogonal Predictors (X\textsubscript{{1}}) and (X\textsubscript{{2}} \textbar X\textsubscript{{1}})  }
   \includegraphics[scale=.45]%
    {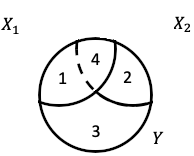}% picture filename
\end{SCfigure}

Not too surprisingly, these Type I sum of squares, when sequentially added, will always equal SS(regression) as calculated by equation (2) or (3), since these additions are always orthogonal. By changing the order in which he placed the predictors in his second computational method, Woolf was able to calculate the “fractions of variance associated with” (1951, p. 113) each of the variables in the final MLR model. In effect, he showed that the sum of squares associated with any predictor in the MLR model was SS(that predictor \textbar all the other predictors in the model), which is perhaps the first recognizable presentation of the simultaneous estimation of the regression sum of squares, Type III, for the regression application. Thus, for a two-predictor situation such as the data in the Appendix, the sum of squares associated with the MLR model is

\begin{equation} SS({\textup{X}_1},{\textup{X}_2)}=SS({\textup{X}_1}|{\textup{X}_2)}+SS({\textup{X}_2}|{\textup{X}_1)}  \end{equation}
Now this result is surprising. The sum of the sequential orthogonal predictors, equation (4), cannot possibly equal the sum of the simultaneous partial predictors, equation (6), in any situation except when the predictors are uncorrelated. This surprising result was also noted by Woolf, “Until now analysis of variance (associated with the regression model)\footnote{The parenthetical addition is included by the authors.}  has given sensible results. But pushed too far, the method soon leads to apparent absurdity” (1951, p. 112).  As a consequence, application of equation (1) for Woolf's data resulted in only 51 \% of the SS(total) actually being accounted for. His conclusion was decidedly disconcerting in 1951 and unfortunately is still disconcerting today. As Woolf concluded, 

\begin{quote}
“There seems to be an enormous deficit in the balance sheet. The paradox, of course, is easily 
	resolved. It is essential for the validity of an analysis of variance that the various components
 	shall be uncorrelated or independent.” (1951, p. 113)
\end{quote} 
	
The actual sum of squares associated with the MLR model, SS(X\textsubscript{1},X\textsubscript{2}), can be calculated in several different ways. A few are presented below. The easiest might be to simply calculate the Extra Sum of Squares contributions from different starting points, such as Woolf did. For our two predictor situation, using the data in Table 1 produces
     
     \begin{equation} SS({\textup{X}_2}|{\textup{X}_1)}=SS(regression)-SS({\textup{X}_1})=24,015.28-18,299.78= 5,715.51 \end{equation}
     
     \begin{equation} SS({\textup{X}_1}|{\textup{X}_2)}=SS(regression)-SS({\textup{X}_2})=24,015.28- 23,371.81 = 643.48  \end{equation} and thus the actual sum of squares associated with the MLR model is

\begin{equation} SS({\textup{X}_1},{\textup{X}_2)}=SS({\textup{X}_1}|{\textup{X}_2)}+SS({\textup{X}_2}|{\textup{X}_1)}=5,715.51 + 643.48 = 6,358.99   \end{equation} Hence, the calculation of SS (regression) as shown on page 2 and in Table 1 does not equal SS(X1, X2), and the answer to the question we posed is no. The sum of squares for the actual MLR model does not equal those produced by the common calculation (i.e., equations 1, 2, 3). This result is perhaps better explained through our second method of calculating the exact sum of squares associated with the MLR model. 

The second method is to first calculate the real data being used in the multiple regression. The coefficients in the MLR model represent the contribution of a particular predictor holding the other predictors constant. In other words, the real data associated with the first predictor is not X\textsubscript{1}, but rather X\textsubscript{1}\textbar X\textsubscript{2} = X\textsubscript{{1.2}}. These values are easily seen as the residuals from the regression

\begin{equation} {\textup{\^{X}}_1}= constant + {\textup{{a}}_2}{\textup{{X}}_2}
\end{equation}

and the values for X\textsubscript{2} \textbar X\textsubscript{1} = X\textsubscript{{2.1}} are the residuals from

\begin{equation} {\textup{\^{X}}_2}= constant + {\textup{{a}}_1}{\textup{{X}}_1}
\end{equation}. If these are the correct values, then the simple regressions of these residualized predictors should reproduce the Extra Sum of Squares presented in equations (7) and (8), and should reproduce the MLR regression coefficients presented in Table 1. These simple regressions appear in Table 3 and confirm these expectations. The result of this reassessment of the actual sum of squares associated with the MLR model is that for the data of Table 1, the 
 \begin{equation} SS(\text{actual total}) = SS({\textup{X}_1},{\textup{X}_2)}+SS(residual)  = 6,358.99 + 2,180.93 = 8,839.92 \end{equation} This leaves 26,196.21 – 8,839.92 = 17,352.29 or 66.2 \% missing from the total sum of squares! 
 
 As seen in Figure 3, only regions associated with the residualized predictors, regions 1 and 2, are included in Y. Region 4 has been entirely removed from X\textsubscript{1}, X\textsubscript{2}, and Y. To quote Woolf, “There seems to be an enormous deficit in the balance sheet.” (1951, p. 113). Unfortunately, unlike the pooling protocol in analysis of variance, MLR does not place the interactive component, region 4, in the residual when it is not included in the model. This confirms the results in Figure 3: only the regions associated with the residualized predictors and the residual are included in Y. Perhaps even more disconcerting than the finding that the sum of squares associated with the MLR model is incorrectly determined, is this finding that the application of the MLR model does not reproduce the total variation of Y.
 
\begin{SCfigure}
  \centering
  \caption{Variation of Y Explained by the Partial Predictors (X\textsubscript{{1.2}}) and (X\textsubscript{{2.1}})}
 \includegraphics[scale=.45]{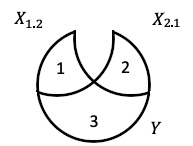} 
\end{SCfigure}
 
Are there other ramifications associated with the classical development of the MLR model? For instance, even though the simple regressions of Table 3 have replicated the unstandardized MLR coefficients, they have failed to produce the same standardized coefficients. Which, if either, are correct? The detailed calculation and explanation of the findings below can be found in (Baird and Bieber, 2020) and hence only the conclusions are illustrated here. To begin, there are two basic principles in regression that will hold for all regression models in which equation (1) is valid. The first is 

\begin{equation} 
\sum_{i=1}^{p} {\textup{z}_i^2} = {\textup{R}^2}
\end{equation}
where {z\textsubscript{{i}}} is the ith standardized MLR coefficient. Several authors, such as Darlington (1968), have recognized that equation (13) holds for data situations in which the predictor variables are uncorrelated. Unfortunately, a simplistic, additive, alternative to equation (13) has not been found for the situation in which the predictors are correlated. However, rather than being the result that the predictors are correlated, perhaps this inability has been the result of using the incorrect data?

For the MLR model of the data in Table 1, if the sum of squares associated with the model is wrong, then what is the correct R\textsuperscript{2}? This is easily calculated, within the context of the original problem, as SS(X\textsubscript{1},X\textsubscript{2})/SS(total) = 6,358.99/26,196.21 = .243. Applying equation (13) to the standardized coefficients in Table 1 produces (.748)\textsuperscript{2} + (.251)\textsuperscript{2} = .623, which is neither the actual R\textsuperscript{2} nor the one presented in Table 1. However, applying equation (13) to the standardized coefficients in Table 3 produces (.467)\textsuperscript{2} + (.157)\textsuperscript{2} = .243, the correct R\textsuperscript{2}. Therefore, the standardized coefficients in Table 3 are logically correct. The use of the correct data which produces the correct standardized coefficients validates the use of equation (13) for all MLR situations regardless of whether a relationship exists within the predictor variables or not. The second principle is the following:

\begin{equation} 
\sum_{i=1}^{p} {\textup{t}_i^2}/p = {f}
\end{equation}. Similarly, for the MLR model of the data in Table 1, what is the correct f? Once again, this is easily calculated as MS(X\textsubscript{1},X\textsubscript{2}) / MS(residual) = [(6,358.99/2)/(2,180.93/18)] = 3,179.50/121.16 = 26.64, is the correct value, not the value for f presented in Table 1. Applying equation (14) to the t values presented in Table 1 produces [(6.87)\textsuperscript{2} + (2.31)\textsuperscript{2}]/2 = 26.24, the correct value (see Baird and Bieber, 2016; 2020). Thus, it can be seen that the application of equation (2) in the MLR situation produces the incorrect estimate of the regression sum of squares since it, as seen above, used the wrong data. The residualized predictors, X\textsubscript{1.2} and X\textsubscript{2.1}, from equations (10) and (11), should be used rather than the original predictors X\textsubscript{1} and X\textsubscript{2}. If equation (2) is modified as the sum of the coefficients times the sum of the cross-products between the residualized predictors and Y, then the correct result is obtained. Using the unstandardized coefficients from Tables 2 and 4 and the Sum of Cross-Products from Table 2 gives

\begin{equation} 
\ {\textup{b}_1}\sum{\textup{X}_{1.2}}Y + {\textup{b}_2}\sum{\textup{X}_{2.1}}Y= (1.455)(3,929.37) + (9.366)(68.71) = 6,358.99 \end{equation}

the value given in equation (9).

It can now be seen that the appropriate model associated with the Extra Sum of Squares presentation use the orthogonal predictors X\textsubscript{1} and X\textsubscript{{2.1}}, or X\textsubscript{2} and X\textsubscript{{1.2}}, rather than the original data. This can be seen in Table 4. These orthogonal functions replicate perfectly the Extra Sum of Squares presentation and are in fact the orthogonal functions presented in Woolf as his solution to the paradox. They are his “formulae for the residuals.”  (1951, p. 114) The application of equation (2) to calculate SS(regression) is validated by either of the two orthogonal regressions in Table 4. 
    
In conclusion, Table 1 provides the traditional summary for an MLR model. As seen above, when the predictors are correlated, only the unstandardized coefficients, the t-values, and the SS(residual) of this table are correct. Unfortunately, the SS(regression), the R\textsuperscript{2}, the f, and the standardized coefficients are all incorrect. It is also unfortunate that the discussion of the Extra Sum of Squares, Type I determination of component sum of squares, is linked inappropriately with the Type III MLR model. As a consequence of these errors, both SPSS (IBM SPSS Statistics, Armonk, NY) and SAS (SAS Software, Cary, NC) use the traditional multiple regression presentation to develop their analysis of variance tables, identical to our Table 1. In an interesting twist, base R (R Foundation for Statistical Computing, Vienna, Austria) follows the Extra Sum of Squares presentation and provides a Type I analysis of variance table, but the parameter estimates from the MLR Type III model.

\newpage

\section{References}

\begin{enumerate}[leftmargin=!,labelindent=5pt,itemindent=-15pt]

\item Baird, G. L., \& Bieber, S. L. (2016). The Goldilocks dilemma: Impacts of multicollinearity-a comparison of simple linear regression, multiple regression, and ordered variable regression models. {\textit{Journal of Modern Applied Statistical Methods, 15(1),}} 18.

\item Baird, G. L., \& Bieber, S. L. (2020). Sampling the Porridge: A Comparison of Ordered Variable Regression with F and R 2 and Multiple Linear Regression with Corrected F and R 2 in the Presence of Multicollinearity.  {\textit{Journal of Modern Applied Statistical Methods, 18(1)}}, 11.  

\item Darlington, R. B. (1968). Multiple regression in psychological research and practice. {\textit{Psychological bulletin, 69(3), 161.}}

\item Fox, J., Weisberg, S., Adler, D., Bates, D., Baud-Bovy, G., Ellison, S., Firth, D., Friendly, M., Gorjanc, G., Graves, S. \& Heiberger, R., {\textit{Package ‘car’}}. (2022). Version R package version 4.1.2. https://cran.r-project.org/web/packages/car/index.html  
                  
\item French, J., {\textit{Package ‘api2lm’}}. (2022). Version R package version 4.1.2. https://cran.r-project.org/web/packages/api2lm/api2lm.pdf        
 
\item Kutner, M. H., Nachtsheim, C. J., Neter, J., \& Li, W. (2005).{\textit{Applied Linear Statistical Models}}, McGraw Hill Irwin, New York. NY, 409.

\item Revelle, W., \& Revelle, M. W. (2015). {\textit{Package ‘psych’}}. The comprehensive R archive network, 337, 338.
   
\item Woolf, B. (1951). Computation and interpretation of multiple regressions. {\textit{Journal of the Royal Statistical Society: Series B (Methodological), 13(1),}} 100-119.

\item Yates, F. (1938). Orthogonal functions and tests of significance in the analysis of variance. {\textit{Supplement to the Journal of the Royal Statistical Society, 5(2),}} 177-180.
\end{enumerate}

\newpage

\begin{table}
 \begin{threeparttable}
\centering
\caption {\textbf{Simple and Multiple Linear Regression Results for Y, X\textsubscript{1} , X\textsubscript{2}}} \label{tab:title} 
\begin{tabular}{llllllllllll}
\firsthline
\cline{1-12}
           & Simple (X\textsubscript{1}) &        &       &  & Simple (X\textsubscript{2}) &       &      &  & MLR (X\textsubscript{1},X\textsubscript{2}) &       &      \\
Source     & SS          & F      & R\textsuperscript{2}    &  & SS          & F     & R\textsuperscript{2}   &  & SS          & F     & R\textsuperscript{2}   \\\hline

Reg & 23,371.81   & 157.22 & .892  &  & 18,299.78   & 44.03 & .699 &  & 24,015.28   & 99.10 & .917 \\
Res   & 2,824.40    &        &       &  & 7,896.43    &       &      &  & 2,180.93    &       &      \\
Total      & 26,196.21   &        &       &  & 26,196.21   &       &      &  & 26,196.21   &       &      \\
           & b           & z      & t     &  & b           & z     & t    &  & b           & z     & t    \\
X\textsubscript{1}         & 1.836       & .945   & 12.54 &  &             &       &      &  & 1.455       & .748  & 6.87 \\
X\textsubscript{2}         &             &        &       &  & 31.173      & .836  & 6.64 &  & 9.366       & .251  & 2.31\\
\lasthline
\end{tabular}
\begin{tablenotes}
      \small
      \item Note: {z} is the standardized coefficient 
    \end{tablenotes}
  \end{threeparttable}
\end{table}

\begin{table}[p]
\centering
\caption {\textbf{Sum of Squares and Cross-Products\\ 
(Mean Centered)}}\label{tab:title)}
\begin{tabular}{llll}
\firsthline
\cline{1-4}
     & Y         & X\textsubscript{1}       & X\textsubscript{2}    \\
\hline
Y    & 26,196.21 &          &       \\
X\textsubscript{1}   & 12,730.59 & 6,934.33 &       \\
X\textsubscript{2}   & 587.04    & 282.33   & 18.83 \\
X\textsubscript{{1.2}} & 3,929.37  &          &       \\
X\textsubscript{{2.1}} & 68.71     &          &      \\
\lasthline
\end{tabular}
\end{table}

\begin{table}[p]
 \begin{threeparttable}
\centering
\caption {\textbf{Simple Regression Results for Y, X\textsubscript{{1.2}}, X\textsubscript{{2.1}}}}\label{tab:title}
\begin{tabular}{llllllllll}
\firsthline
\cline{1-8}
       & Simple (X\textsubscript{{1.2}}) &           &        &   & Simple (X\textsubscript{{2.1}}) 
      \\Source     & SS            & f         & R\textsuperscript{2}    &  & SS            & f    & R\textsuperscript{2}   \\
\hline
Regression & 5,715.51      & 5.30      & .218 &  & 643.48        & .48  & .025 \\
Residual   & 20,480.71     &   &      &  & 25,552.73               &      &      \\
Total      & 26,196.21     &  &      &   & 26,196.21               &      &      \\
\hline
           & b             & z         & t    &  & b             & z    & t    \\
\hline
X\textsubscript{{1.2}}       & 1.455         & .467      & 2.30 &  &               &      &      \\
X\textsubscript{{2.1}}       &               &           &      &  & 9.366         & .157 & .69  \\
\hline
\end{tabular}
\begin{tablenotes}
      \small
      \item Note: {z} is the standardized coefficient 
    \end{tablenotes}
  \end{threeparttable}
\end{table}

\clearpage

\begin{table}[p]
 \begin{threeparttable}
\centering
\caption {\textbf{Orthogonal Function Regression Results: Y, X\textsubscript{{1}}, X\textsubscript{{2.1}} and Y, X\textsubscript{{2}}, X\textsubscript{{1.2}}}} \label{tab:title}
\begin{tabular}{llllllll}
\firsthline
\cline{1-8}
           & X\textsubscript{{1}} and X\textsubscript{{2.1}} &           &       &  & X\textsubscript{{2}} and X\textsubscript{{1.2}} &       &      \\
Source     & SS          & f         &  R\textsuperscript{2}    &  & SS          & f     &  R\textsuperscript{2}   \\
\hline
Regression & 24,015.28   & 99.10     & .917  &  & 24,015.28   & 99.10 & .917 \\
Residual   & 2,180.93    &   &       &  &     2,180.93        &       &      \\
Total      & 26,196.21   &  &       &  &     26,196.21        &       &      \\
\hline
           & b           & z         & t     &  & b           & z     & t    \\
\hline
X\textsubscript{{1}}        & 1.836       & .945      & 13.89 &  &             &       &      \\
X\textsubscript{{2.1}}       & 9.366       & .157      & 2.31  &  &             &       &      \\
X\textsubscript{{2}}         &             &           &       &  & 31.173      & .836  &    12.29  \\
X\textsubscript{{1.2}}       &             &           &       &  & 1.455       & .467  &     6.87\\
\hline
\end{tabular}
\begin{tablenotes}
      \small
      \item Note: {z} is the standardized coefficient 
    \end{tablenotes}
  \end{threeparttable}
\end{table}

\clearpage

\begin{singlespace}

\section{Appendix A}

\begin{verbatim}


#R Code 

#Dataset from Kutner, et al. (2005) : 
library("api2lm")

X1=dwaine$targetpop
X2=dwaine$dpi
Y= dwaine$sales
###Table 1

## SLR Models
SLR1 = lm(Y~ X1)

#ANOVA 
anova(SLR1) 

#Coefficients 
summary(SLR1) 
B1= (1.8359*(sd(X1))/sd(Y))
print(B1)

SLR2 = lm(Y~ X2)

#ANOVA
anova(SLR2) 

#Coefficients 
summary(SLR2)
B2= (31.173*(sd(X2))/sd(Y))
print(B2)

### MLR Model
MLR = lm(Y~ X1 + X2)

#ANOVA
library("car")
Anova(MLR, type="III") 
anova(MLR) #type I

#Coefficients 
summary(MLR) 
B1= (1.4546*(sd(X1))/sd(Y))
print(B1)
B2= (9.3655*(sd(X2))/sd(Y))
print(B2)


###Table 3

##Residualized Predictors
residX1 = lm(X1~X2)
X1.2=resid(residX1)

#SLR Models
SLR1.2 = lm(Y~ X1.2)
summary(SLR1.2)
anova(SLR1.2)

residX2 = lm(X2~X1)
X2.1=resid(residX2)

SLR2.1 = lm(Y~ X2.1)
summary(SLR2.1)
anova(SLR2.1)

#Correct Standardized Coefficients and Correct R2
B1.2= (1.4546*(sd(X1.2))/sd(Y))
print(B1.2)
Correct_R2=(B1.2^2)
print(Correct_R2)

B2.1= (9.3655*(sd(X2.1))/sd(Y))
print(B2.1)
Correct_R2=(B2.1^2)
print(Correct_R2)

###Table 2 SSCP
library("psych")
all=data.frame(Y,X1,X2,X1.2,X2.1)
sscp=cov(all)/.05
print(sscp)


###Table 4
##Orthogonal Functions Model
OFM1 = lm(Y~ X1+X2.1)
anova(OFM1)

#Coefficients and R2
summary(OFM1)

B1= (1.8359*(sd(X1))/sd(Y))
print(B1)

B2.1= (9.3655*(sd(X2.1))/sd(Y))
print(B2.1)

R2=(B1^2)+(B2.1^2)
print(R2)

###Orthogonal Functions Model 2
OFM2 = lm(Y~ X2+X1.2)
anova(OFM2)

##Coefficients ND R2
summary(OFM2)

B2= (31.1732*(sd(X2))/sd(Y))
print(B2)

B1.2= (1.4546*(sd(X1.2))/sd(Y))
print(B1.2)

R2=(B2^2)+(B1.2^2)
print(R2)


###Equation 12
MLR = lm(Y~ X1+X2)

#Type III Sums of Squares
Anova(MLR, type="III") 
Actual_Total_SS=(5715.5+643.5+2180.9)
print(Actual_Total_SS)

###Equation 13
B1..2= (1.4546*(sd(X1.2))/sd(Y))
B2..1= (9.3655*(sd(X2.1))/sd(Y))
Correct_R2=(B1..2^2)+(B2..1^2)
print(Correct_R2)

\end{verbatim}
\end{singlespace}

\end{document}